\documentclass{article}

\PassOptionsToPackage{numbers}{natbib}

\usepackage[preprint]{neurips_2021}

\usepackage[utf8]{inputenc} 
\usepackage[T1]{fontenc}    
\usepackage{hyperref}       
\usepackage{url}            
\usepackage{booktabs}       
\usepackage{amsfonts}       
\usepackage{nicefrac}       
\usepackage{microtype}      
\usepackage{xcolor}         
\usepackage{graphics}

\usepackage{adjustbox}
\usepackage{multirow}
\usepackage{verbatim}

\PassOptionsToPackage{skip=0pt}{figure}

\title{Quantifying and Maximizing the Benefits of Back-End Noise Adaption on Attention-Based Speech Recognition Models}

\author{
Coleman Hooper \\
Harvard University \\
chooper@college.harvard.edu \\
\and
\textbf{Thierry Tambe} \\
Harvard University \\
ttambe@g.harvard.edu \\
\and
\textbf{Gu-Yeon Wei} \\
Harvard University\\
gywei@g.harvard.edu \\

}

\begin{document}

\maketitle

\begin{abstract}
  This work analyzes how attention-based Bidirectional Long Short-Term Memory (BLSTM) models adapt to noise-augmented speech. We identify crucial components for noise adaptation in BLSTM models by freezing model components during fine-tuning. We first freeze larger model subnetworks and then pursue a fine-grained freezing approach in the encoder after identifying its importance for noise adaptation. The first encoder layer is shown to be crucial for noise adaptation, and the weights are shown to be more important than the other layers.  Appreciable accuracy benefits are identified when fine-tuning on a target noisy environment from a model pretrained with noisy speech relative to fine-tuning from a model pretrained with only clean speech when tested on the target noisy environment. For this analysis, we produce our own dataset augmentation tool and it is open-sourced to encourage future efforts in exploring noise adaptation in ASR.

\end{abstract}

\section{Introduction}
\label{intro}
 
Automatic Speech Recognition (ASR) is becoming ubiquitous for controlling household appliances and mobile devices. To maintain accuracy in target deployment environments, ASR models must be robust to environmental noise. There are two typical approaches to reduce the impacts of noise on ASR performance, which are termed front-end and back-end \cite{zhang2018deep}\cite{6732927}. Front-end methods aim to reduce the amount of noise in speech such that ASR models can decode it in a similar manner to clean speech. Back-end methods instead aim to adapt the model itself to be robust to noisy speech. These methods either fine-tune the entire model with noise \cite{conf/interspeech/MirsamadiH15} or adapt specific model components to the target noisy environment \cite{michel-neubig-2018-extreme}. Noise can be injected into the dataset prior to training \cite{noisytrainingforDNNs} or added to clean speech on-the-fly during training \cite{9054130}. Training with noise has been shown to improve both ASR model performance and generalizability \cite{noisytrainingforDNNs}. Transfer learning has been applied widely across languages or across accents in ASR  \cite{DBLP:journals/corr/KunzeKKKJS17}\cite{6855089}. Previous studies have analyzed how cross-domain transfer learning improves end task performance \cite{su-etal-2012-translation}, assessed the contribution of specific weights during transfer learning \cite{vilar-2018-learning}, and shown that earlier model layers tend to be more general and later model layers tend to be more specific for the target task \cite{howtransferable}. 

\begin{figure}
  \centering
  \includegraphics[width=0.6\linewidth]{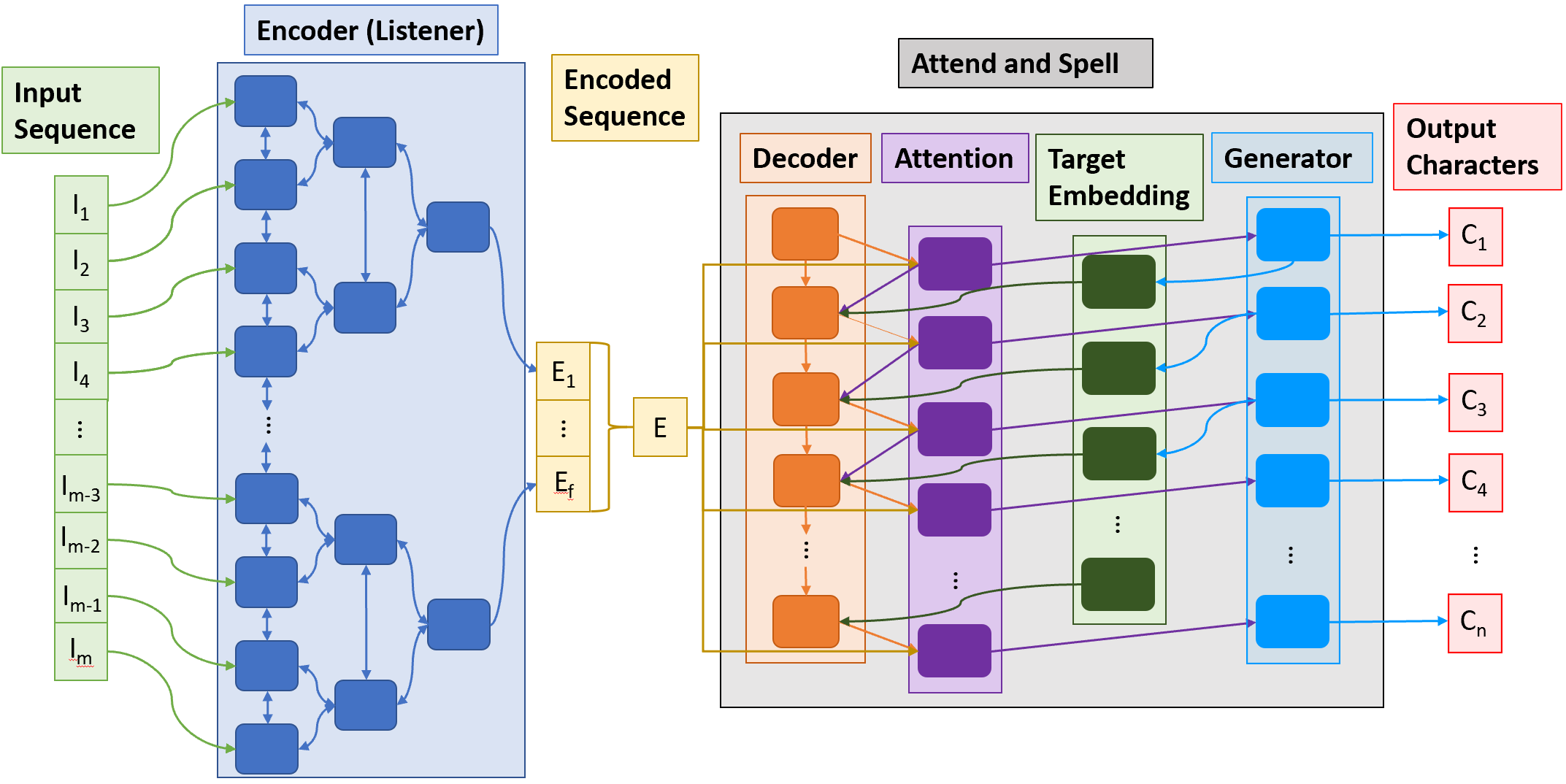}
  \caption{Diagram showing the Listen, Attend, and Spell architecture from \cite{DBLP:journals/corr/ChanJLV15}.}
      \label{fig:attentionBLSTM}
      \vspace{-1.5em}
\end{figure}

Freezing subnetworks has been found to be a useful technique for identifying crucial model components for adapting to the end task in machine translation \cite{Thompson_2018}. Additionally, freezing individual layers can identify which layers must be adapted to the end task and which can be left unchanged during transfer learning \cite{howtransferable}. Freezing model weights allows for efficient hardware deployment, as frozen weights can be securely stored and kept on-chip and do not need to be reloaded from DRAM, saving considerable latency and energy. Additionally, if only a small proportion of weights must be fine-tuned, these weights can either be fine-tuned on-device or changed depending on the noisy environment while leaving the frozen parameters unchanged. Finally, freezing parameters will improve training times since fewer weights have to be updated during training. 

Among the most prominent model architectures for ASR are attention-based sequence-to-sequence models \cite{DBLP:journals/corr/ChanJLV15}\cite{8462105}\cite{Zeyer_2018}\cite{Irie_2019}. This work assesses the impacts of noise on attention-based BLSTM models by freezing model components during fine-tuning on noisy speech. We used the Listen, Attend, and Spell (LAS) model architecture for our experiments \cite{DBLP:journals/corr/ChanJLV15}. LAS is an attention-based BLSTM model architecture that has a pyramidal encoder (the Listener) followed by an attention-based decoder (the Attend and Spell portion of the architecture). Figure \ref{fig:attentionBLSTM} provides a diagram of this architecture. 
We first assess the impacts of freezing larger model subnetworks during fine-tuning and identify the encoder as the crucial subnetwork for noise adaptation. We then investigate fine-grained freezing strategies to maximize the number of parameters that can be frozen without significant accuracy degradation, thereby identifying the weights in the first encoder layer as being particularly important for noise adaptation. We observe distinct accuracy benefits when pretraining attention-based BLSTM models on generic noise compared to pretraining these models with only clean speech prior to adapting the model on a target noisy environment, especially when freezing model components during fine-tuning. Additionally, we demonstrate the benefits of using diverse room environments and noise types for pretraining instead of only augmenting speech with additive noise. We constructed and open-sourced a dataset augmentation tool that allows the user to produce synthetic noisy datasets for a desired room configuration using available clean speech and background noise datasets.\footnote{Code is available at \url{https://github.com/chooper1/NoiseAugment.}}

\section{Experimental Setup}

\subsection{Dataset Preparation}
\label{datasets}

\begin{figure}
  \centering
  \includegraphics[width=0.6\linewidth]{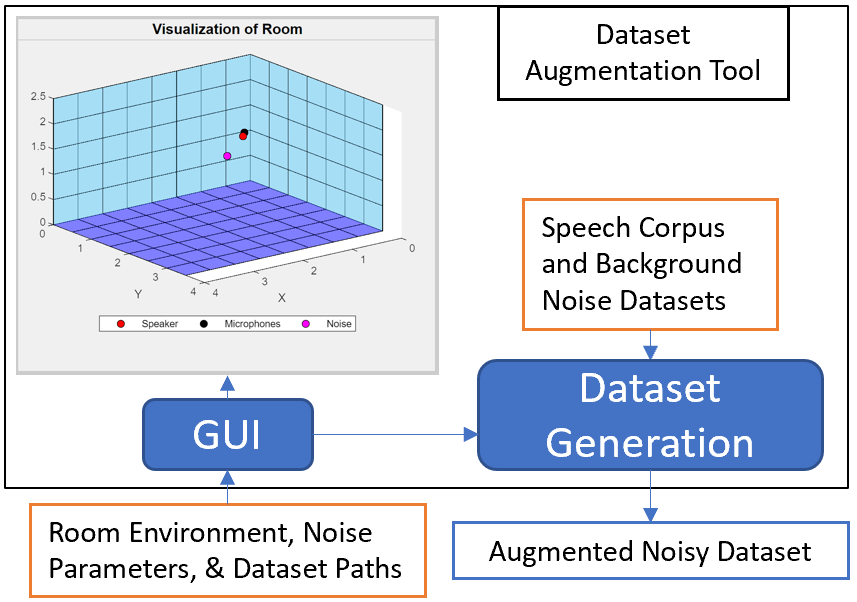}
  \caption{Diagram showing how noisy datasets were produced using NoiseAugment. The room environment visualization is updated in real time as noise parameters are changed in the GUI.}
      \label{fig:datasetpipeline}
      \vspace{-1.5em}
\end{figure}

We constructed NoiseAugment, a dataset augmentation tool in Matlab which was built on top of the Roomsimove Matlab toolbox \cite{roomsimove}. Our tool allows for several audio sound sources (a desired speaker plus one or more noise sources) to be placed at different positions in a simulated room environment with reconfigurable dimensions and microphone positions. For each input speech sample, the clean speech and background noise samples are played in the simulated room and recorded by the microphone(s), using the Roomsimove toolbox to compute the room impulse response. The tool also includes an interactive GUI that allows the user to visualize the positions of the speaker, noise sources, and microphones. The tool allows for automated dataset generation using publicly available speech and background noise datasets. Additional settings allow the user to set the probability of each noise source being included, as well as the amplitude of each noise source relative to the amplitude of the speaker. Although this tool does not allow the user to accurately model a desired target room as it does not allow for objects to be placed within the room, it allows the user to augment datasets with simulated reverberation in order to assess how ASR models perform across a diverse range of noisy reverberant environments. A diagram outlining our dataset generation pipeline is shown in Figure \ref{fig:datasetpipeline}. 

We used the LibriSpeech dataset \cite{librispeech} for clean speech samples, and then produced augmented datasets using our noise augmentation tool along with publicly available noise datasets. We included both the train-clean-100 and train-clean-360 splits in the LibriSpeech training set. The amplitude of the noise was scaled to be between 0.2 and 0.4 times the amplitude of the speaker. A single microphone was placed in the simulated room in all augmented datasets, and the background noise datasets were all segmented into training, validation, and test splits. We produced datasets for three different room environments; a living room, a cocktail party, and a kitchen. The "Home" (living room) dataset included household noise from the CHiME-Home \cite{chime-home} dataset, as well as samples from  Urbansound8K \cite{urbansound8k} as urban noise coming in through a window. The "Cocktail" (cocktail party) dataset used additional randomly selected samples from the LibriSpeech dataset as background noise. Note that we ensured that the desired speaker was different than the other speakers that were included as noise. The "Kitchen" dataset included appliance noises and other kitchen noises from the ESC-50 \cite{DVN/YDEPUT_2015}, NAR \cite{janvier:hal-00952092}, and Freiburg 106 \cite{freiburg} datasets.

We produced a "Room" baseline dataset which contained a mix of augmented samples from 5 other simulated room environments, as well as a "No Room" baseline dataset which only augmented speech with additive noise (without using a synthetic room environment) and did not expose the model to varied noise types and room environments during training. The Room baseline allowed the model to encounter different reverberant room environments and noise types in order to improve its performance on new room environments. The Room baseline used the Clotho \cite{clotho} and WHAM! \cite{wham!} datasets, and the No Room baseline was augmented with the WHAM! \cite{wham!} dataset.

\subsection{Transfer Learning and Freezing Model Components}
\label{tl_and_freeze}
We trained and evaluated LAS models using the OpenNMT-py framework \cite{opennmt}. We modified this framework to support freezing model components. We used models with 4 hidden layers in the encoder and 1 hidden layer in the decoder, each with hidden layer sizes of 512. We used pooling factors of 1, 1, 2, and 2 in the four encoder layers, which reduced the number of timesteps in the final two encoder layers by factors of 2. We trained all models with a batch size of 32, a learning rate of 0.0003, and a learning rate decay factor of 0.95. ADAM \cite{adam} was used as the optimizer. Experiments were performed on the FASRC Odyssey cluster, which is supported by the FAS Division of Science Research Computing Group at Harvard University. The LAS model had 9.83 million parameters, and the number of parameters in each subnetwork is given in Table \ref{tab:numparams512}. We trained the clean baseline model on the clean LibriSpeech dataset and the noisy baseline models on the Room and No Room datasets for 400 000 steps. These noisy baseline models performed poorly, so we also trained noisy baseline models in two stages by first training for 350 000 steps on the clean LibriSpeech dataset and then for 50 000 steps on the Room and No Room datasets. When fine-tuning on a target noisy environment, we started from a pretrained model and trained the model to convergence for an additional 50 000 steps on a specific noisy dataset. We used the Word Error Rate (WER) as the metric for assessing ASR performance. 

\begin{table}[th]
\caption{Number of parameters in each model subnetwork.}
\label{tab:numparams512}
\begin{adjustbox}{width=0.5\columnwidth,center}
  \centering
  \begin{tabular}{c c c c c c}
     \toprule
     Encoder & Decoder & Attention & Embedding & Generator  \\
    \midrule
     5.85M & 3.16M & 786K & 16K & 16K    \\
    \bottomrule
  \end{tabular}
\end{adjustbox}
\end{table}

To investigate the importance of each subnetwork for noise adaptation, we fine-tuned both with the specific subnetwork frozen and with every other subnetwork frozen, which mirrored the subnetwork freezing experiments from \cite{Thompson_2018}. We performed these experiments using the clean baseline and the Room and No Room baselines as starting points. After identifying the encoder as crucial for noise adaptation, we assessed the importance of the projection layer, layer normalization layers, weights, and biases by fine-tuning models both with the specific layers frozen and with all other layers frozen. We then assessed the importance of weights and biases in each encoder layer by freezing all other layers while fine-tuning. Finally, we individually assessed the importance of the forward input, forward hidden state, reverse input, and reverse hidden state weights and biases by fine-tuning the weights and biases of one type while leaving the rest frozen.

\section{Results}

\subsection{Transfer Learning}

\begin{table}[th]
\caption{Cross-Dataset WER results with models pretrained with clean speech and fine-tuned on a target noisy environment. The lowest WER for each test set is bolded.}
   \label{tab:crossdatasetfromclean}
  \centering
  \begin{adjustbox}{width=0.7\columnwidth,center}
  \begin{tabular}{ c c c c c c c c c c c c c c }
    \toprule
\multirow{2}{*}{\textbf{Fine-Tuning Dataset}} & \multicolumn{6}{c}{\textbf{Test Dataset}}                   \\ 
\cmidrule(l{0em}r{0em}){2-7}
& Clean  & Home & Cocktail & Kitchen & No Room & Room                   \\ \midrule
Clean Baseline                  & \textbf{6.23}\% & 30.15\% & 40.70\% & 24.50\% & 16.13\% & 29.89\%  \\ \midrule
Home                   & 7.41\% & \textbf{17.06}\% & 37.53\% & 16.09\% & \textbf{12.17}\% & 22.80\%  \\ \midrule
Cocktail                & 8.15\% & 24.78\% & \textbf{19.17}\% & 23.17\% & 12.33\% & \textbf{18.88}\%  \\ \midrule
Kitchen               & 6.76\% & 20.99\% & 36.52\% & \textbf{15.27}\% & 12.88\% & 24.59\%  \\ 

\bottomrule
  \end{tabular}
  \end{adjustbox}
\end{table}

Table \ref{tab:crossdatasetfromclean} provides cross-dataset results for models fine-tuned from the clean baseline model. Fine-tuning on the target noisy dataset led to a large reduction in the WER relative to the clean baseline when tested on the same noisy environment. However, models fine-tuned on one specific noisy environment did not generalize well to other noisy environments; for example, fine-tuning on the Kitchen and Home datasets did not significantly reduce the WER when tested on the Cocktail dataset. 

\begin{table}[th]

\caption{Cross-Dataset WER results for the baseline models. The first two models were trained using only noisy data, and the second two models were trained in two stages (first using clean data and then using noisy data). The lowest WER for each test set is bolded. }
   \label{tab:crossdatasetbaseline}
  \centering
  \begin{adjustbox}{width=0.7\columnwidth,center}
  \begin{tabular}{ c c c c c c c c c c c c c c }
    \toprule
\multirow{2}{*}{\textbf{Training Dataset}} & \multicolumn{6}{c}{\textbf{Test Dataset}}                   \\ 
\cmidrule(l{0em}r{0em}){2-7}
& Clean  & Home & Cocktail & Kitchen & No Room & Room                   \\ \midrule

No Room Only             & 10.34\% & 26.75\% & 42.90\% & 22.48\% & 12.08\% & 27.50\%  \\ \midrule
Room Only                & 16.34\% & 28.13\% & 28.25\% & 25.74\% & 19.19\% & 25.28\%  \\ \midrule

Clean then No Room             & \textbf{7.00}\% & 24.23\% & 37.23\% & 21.91\% & 10.86\% & 24.23\%  \\ \midrule
Clean then Room                & 7.50\% & \textbf{19.54}\% & \textbf{19.61}\% & \textbf{17.79}\% & \textbf{10.52}\% & \textbf{16.19}\%  \\  

\bottomrule
  \end{tabular}
  \end{adjustbox}
\end{table}

Table \ref{tab:crossdatasetbaseline} shows the cross-dataset results for the baseline models trained with noise. The models pretrained with only noisy speech ("No Room Only" and "Room Only") performed poorly across all test sets. Conversely, the models which were pretrained in two stages ("Clean then No Room" and "Clean then Room") were able to generalize well to each specific noisy dataset. The models pretrained in two stages were used as the starting point for our fine-tuning experiments. Pretraining with the No Room baseline resulted in WER reductions of 5.92\%, 3.47\%, and 2.59\% on the Home, Cocktail, and Kitchen datasets, respectively, compared with the baseline model trained with only clean speech. Pretraining with the Room baseline resulted in WER reductions of 10.61\%, 21.09\%, and 6.71\% on the Home, Cocktail, and Kitchen datasets, respectively. These results show that a model pretrained with a varied set of noisy environments can generalize well to unseen environments. 

\begin{table}[th]
 \caption{Cross-Dataset WER with models pretrained in two stages (with clean speech and then with noisy speech) and then fine-tuned with specific noise. The lowest WER for each test set is bolded. }
   \label{tab:crossdatasetfromnoise}
  \centering
  \begin{adjustbox}{width=0.85\columnwidth,center}
  \begin{tabular}{ c c c c c c c c c c c c c c c }
    \toprule
\textbf{Pretraining Dataset} & \multirow{2}{*}{\textbf{Fine-Tuning Dataset}} & \multicolumn{6}{c}{\textbf{Test Dataset}}                   \\ 
\cmidrule(l{0em}r{0em}){3-8} 

\textbf{(Second Stage)} & & Clean  & Home & Cocktail & Kitchen  & No Room & Room                     \\  \midrule

\multirow{3}{*}{No Room} & Home                   & 6.97\% & 16.21\% & 36.09\% & 15.73\% & 10.78\% & 22.07\%  \\ \cmidrule(l{0em}r{0em}){2-8}
& Cocktail              & 7.38\% & 22.61\% & 17.51\% & 20.95\% & 11.39\% & 17.10\%  \\ \cmidrule(l{0em}r{0em}){2-8}
& Kitchen                & 6.69\% & 19.20\% & 35.63\% & 14.72\% & 10.48\% & 22.87\%  \\ \midrule
\multirow{3}{*}{Room} & Home                   & \textbf{6.76}\% & \textbf{14.74}\% & 26.79\% & 15.02\% & 10.68\% & 18.03\%  \\ \cmidrule(l{0em}r{0em}){2-8}
& Cocktail         & 7.02\% & 20.40\% & \textbf{16.18}\% & 18.89\% & \textbf{10.18}\% & \textbf{15.58}\%  \\ \cmidrule(l{0em}r{0em}){2-8}
& Kitchen                & 6.87\% & 17.13\% & 27.10\% & \textbf{13.43}\% & 10.47\% & 18.91\%  \\

\bottomrule
  \end{tabular}
  \end{adjustbox}
\end{table}

Table \ref{tab:crossdatasetfromnoise} shows the cross-dataset WER results when fine-tuning from models pretrained with noise. Models fine-tuned on one specific noisy environment from the Room baseline vastly outperformed models fine-tuned from clean speech and from the No Room baseline when tested on other noisy environments, demonstrating that they retain generalizability from pretraining. Additionally, models fine-tuned from a generic noisy baseline to a specific noisy environment performed better on the target noisy environment when compared with the performance of models fine-tuned from the baseline clean model; pretraining with the Room baseline resulted in WER reductions of 2.32\%, 2.99\%, and 1.84\% on the Home, Cocktail, and Kitchen datasets, respectively, relative to pretraining with the clean baseline. This demonstrates the benefits of pretraining with noise and then fine-tuning on the target noisy environment for improving performance on the target environment and generalizability.

\subsection{Freezing Subnetworks}
\label{freezing subnetworks clean}

\begin{figure*}
  \centering
  \includegraphics[width=\textwidth]{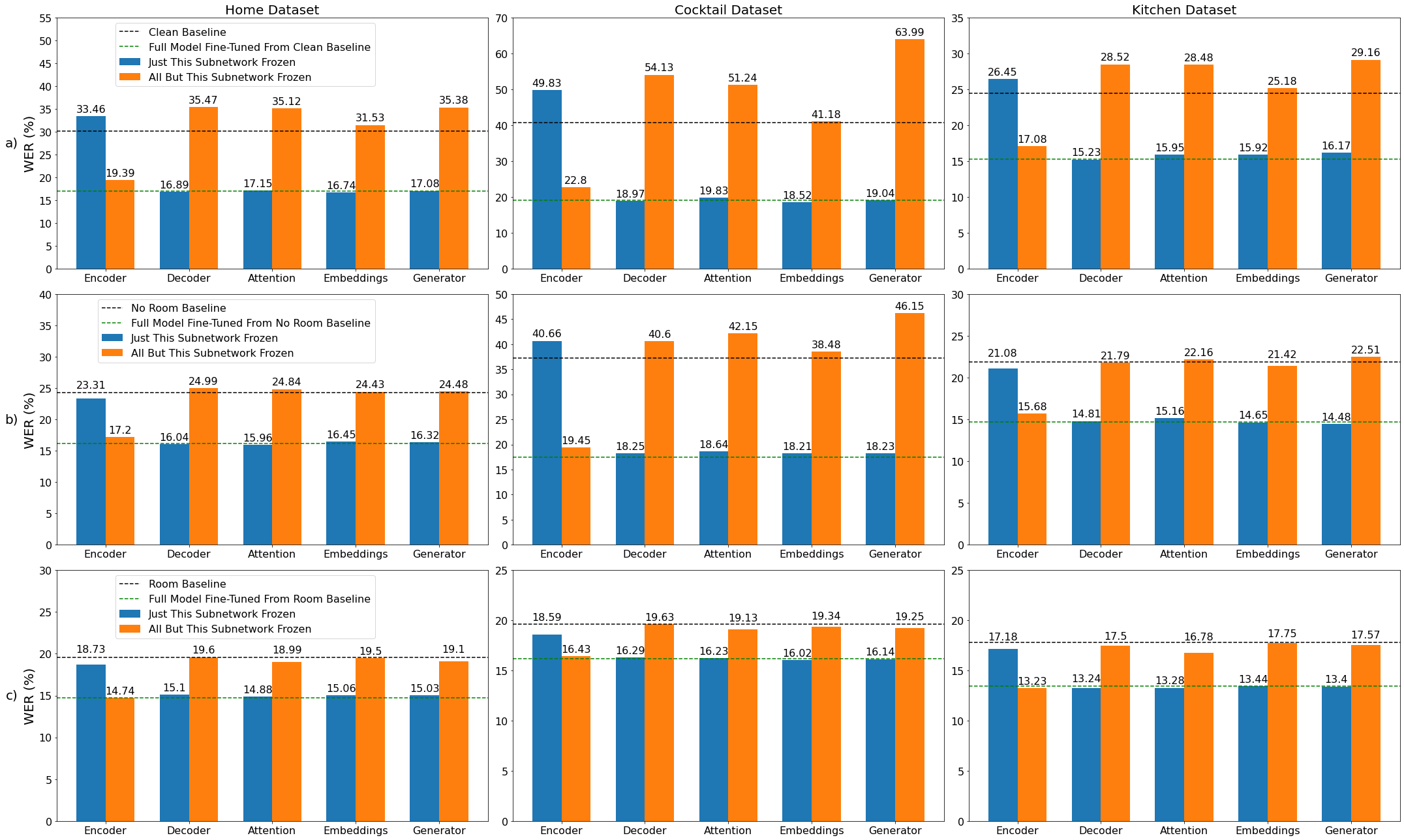}
  \caption{Freezing model components during fine-tuning on a target noisy environment. a) Subnetwork freezing from clean speech. b) Subnetwork freezing from the No Room baseline dataset. c) Subnetwork freezing from the Room baseline dataset.}
      \label{fig:freezingmodelcomponents}
      \vspace{-1.5em}
\end{figure*}

The results from freezing subnetworks while fine-tuning from clean speech are shown in a) in Figure \ref{fig:freezingmodelcomponents}. These results demonstrate that it is crucial to fine-tune the encoder on noise and that the other model subnetworks are unimportant for noise adaptation. Freezing just the decoder, attention layer, embeddings, or generator did not result in a significantly degraded WER relative to the full fine-tuned model, whereas if the encoder was frozen, the WER of the fine-tuned model was worse than the initial model. Fine-tuning with all subnetworks other than the encoder frozen resulted in WER degradation of only 2.33\%, 3.63\%, and 1.81\% on the Home, Cocktail, and Kitchen datasets, respectively. The results from our subnetwork freezing experiments when fine-tuning from generic noisy speech are shown in b) and c) in Figure \ref{fig:freezingmodelcomponents}. When the model has been pretrained using the Room baseline, freezing subnetworks other than the encoder during the fine-tuning process results in minimal accuracy degradation relative
to fine-tuning the entire model. The WER of the models pretrained using the Room dataset and with only the encoder fine-tuned differed from the models with all components fine-tuned by 0\%, 0.25\%, and -0.2\% for the Home, Cocktail, and Kitchen datasets, respectively. 

\subsection{Fine-Grained Encoder Subcomponent Freezing}

\begin{table}[th]
\caption{WER when freezing specific encoder parameters while fine-tuning from the Room dataset. }
   \label{tab:finegrainedfreezing}
  \centering
  \begin{adjustbox}{width=0.85\columnwidth,center}
  \begin{tabular}{ c c c c c c c c c c}
    \toprule
\textbf{Training \&} &  \multicolumn{4}{c}{\textbf{Just Freezing These Layers}} &  \multicolumn{4}{c}{\textbf{Freezing All But These Layers}} \\
\cmidrule(l{0em}r{0.2em}){2-5} \cmidrule(l{0.2em}r{0em}){6-9}
\textbf{Test Dataset} & Projection & Weights & Biases & Layernorm & Projection & Weights & Biases & Layernorm                 \\ \midrule

Home & 14.68\% & 17.36\% & 14.75\% & 15.05\% & 18.71\% & 14.76\% & 18.09\% & 18.85\% \\ \midrule

Cocktail & 16.28\% & 19.38\% & 16.32\% & 16.52\% & 19.66\% & 16.94\% & 19.33\% & 19.36\%  \\ \midrule

Kitchen & 13.38\% & 15.09\% & 13.59\% & 12.99\% & 17.14\% & 13.40\% & 15.70\% & 16.41\% \\

\bottomrule
  \end{tabular}
  \end{adjustbox}
\end{table}

\begin{table}[th]
\caption{WER when freezing individual encoder layers and when freezing input and hidden state weights while fine-tuning from the Room dataset. }
   \label{tab:layerbylayer_ihhh}
  \centering
  \begin{adjustbox}{width=\columnwidth,center}
  \begin{tabular}{ c c c c c c c c c c}
    \toprule
\textbf{Training \&} &  \multicolumn{4}{c}{\textbf{Freezing All But This Layer}} &  \multicolumn{4}{c}{\textbf{Freezing All But These Weights and Biases}} \\
\cmidrule(l{0em}r{0.2em}){2-5} \cmidrule(l{0.2em}r{0em}){6-9}
\textbf{Test Dataset} & Layer 1 & Layer 2 & Layer 3 & Layer 4 & Forward Input & Reverse Input & Forward Hidden & Reverse Hidden             \\ \midrule

Home &  16.27\% & 16.51\% & 18.22\% & 18.56\% & 16.69\%  & 16.51\%  & 16.63\%  & 16.66\% \\ \midrule

Cocktail &  18.28\% & 18.66\% & 18.78\% & 19.04\% & 18.39\%  & 18.18\%  & 18.56\%  & 18.05\% \\ \midrule

Kitchen &  14.49\% & 15.22\% & 16.15\% & 16.31\% & 15.02\%  & 14.79\%  & 14.77\%  & 14.83\% \\

\bottomrule
  \end{tabular}
  \end{adjustbox}
\end{table}

The results from freezing the projection layer, weights, biases, and layer normalization layers are shown in Table \ref{tab:finegrainedfreezing}. The projection layer and layer normalization layers in the encoder were found to not be crucial for adapting to the target noise environment; these layers were therefore frozen in the subsequent fine-grained freezing experiments. The weights in each encoder layer were found to be particularly vital for noise adaptation. The layer-by-layer freezing results when fine-tuning from the Room dataset are given in Table \ref{tab:layerbylayer_ihhh}. The first encoder layer was the most important for noise adaptation across all three datasets. Additionally, each subsequent encoder layer was progressively less important for noise adaptation. The results when freezing the input and hidden state weights and biases in the forward or reverse layers are also given in Table \ref{tab:layerbylayer_ihhh}. The input and hidden state weights and biases across the forward and reverse layers were equally important for noise adaptation. Freezing all model parameters except for the first encoder layer therefore maximizes the number of frozen model parameters while minimizing WER degradation. Fine-tuning only the weights and biases in the first encoder layer led to WERs of 16.27\%, 18.28\%, and 14.49\% on the Home, Cocktail, and Kitchen datasets, respectively; 91.29\% of model weights could therefore be frozen with WER degradation of 1.53\%, 2.10\%, and 1.26\%, respectively, across the Home, Cocktail, and Kitchen datasets, relative to fine-tuning the entire model from the Room dataset. Additionally, the WER when fine-tuning only the weights and biases in the first encoder layer from the Room baseline are 0.79\%, 0.89\%, and 0.78\% lower on the Home, Cocktail, and Kitchen datasets, respectively, compared to fine-tuning the full model from the Clean baseline.

\section{Conclusion}
\label{conclusion}
It is crucial that ASR models are robust to noise to avoid accuracy degradation in noisy environments. In LAS models, the encoder was critical for noise adaptation, with each successive encoder layer being of decreasing importance. By fine-tuning only the weights and biases in the first encoder layer and pretraining the other subnetworks with noise, all but 8.71\% of the model parameters were able to be frozen while achieving WERs of 16.27\%, 18.28\%, and 14.49\% on the Home, Cocktail, and Kitchen datasets, respectively, which are 0.79\%, 0.89\%, and 0.78\% lower across these three datasets relative to fine-tuning the full model from the clean baseline. Adapting only this layer would allow the weights and biases for this layer to either be adapted on-device or changed out depending on the noisy environment. The remaining model weights could be kept on-chip and not reloaded from off-chip DRAM, which would lead to significant energy and latency savings while still having the model be robust to environmental noise. These results help motivate future exploration to identify crucial model components for noise adaptation across other ASR model architectures. 

\medskip

\bibliography{neurips_2021}

\appendix

\end{document}